\documentclass[aps,nobibnotes]{revtex4}

\usepackage{graphicx}

\begin{document}

\title{Competition of fusion and quasi-fission in the reactions leading 
to production of the superheavy elements}

\author{M. Veselsky}

\email{fyzimarv@savba.sk}

\affiliation{Institute of Physics, Slovak Academy of Sciences, Bratislava, Slovakia}




\begin{abstract}
The mechanism of fusion hindrance, an effect observed 
in the reactions of cold, warm and hot fusion leading to production 
of the superheavy elements, is investigated. A systematics 
of transfermium production cross sections is used to determine fusion 
probabilities. Mechanism of fusion hindrance is described as a competition 
of fusion and quasi-fission. Available evaporation residue cross sections 
in the superheavy region are reproduced satisfactorily. Analysis of 
the measured capture cross sections is performed and a sudden disappearance  
of the capture cross sections is observed at low fusion probabilities. 
A dependence of the fusion hindrance on the asymmetry of the 
projectile-target system is investigated using the available data. 
The most promising pathways for further experiments are suggested. 
\end{abstract}

\maketitle

\section*{Introduction}

In the recent years, the heavy elements up to Z=112 
have been synthesized using cold fusion reactions 
with Pb, Bi targets in the evaporation channel with emission 
of one neutron \cite{Z112Cold}. The experimentalists 
had to face a steep decrease of cross sections up to the picobarn 
level due to increasing fusion hindrance whose origin was unclear. 
The same level of cross sections has been 
reached in the hot fusion reactions with emission 
of 3-4 neutrons using $^{48}$Ca beams which lead to 
synthesis of relatively neutron-rich isotopes of elements 112,114 and 116 \cite{Z112Vass,Z114Vass,Z114GNS,Z116GNS}. 
Again the fusion hindrance was observed. The possibility to describe 
fusion hindrance in both cold and hot fusion in a unified way 
as a competition between formation of the compound nucleus and 
a fast fission-like process ( quasi-fission ) was suggested 
in our article \cite{MVSHE} using a simple phenomenological model. 
A comparison of the recent experimental results to the results 
of the model calculation is provided in the present article. 
Furthermore, additional investigations on the nature of the 
fusion process are carried out using available data on capture 
cross section. An additional dynamical fusion hindrance 
is predicted based on available experimental evaporation residue 
data from the reactions where heavy nuclei are produced in the 
symmetric projectile-target combinations approaching the asymmetry 
of the fission channel. 

\section*{Statistical model for competition of fusion and quasi-fission}

In our previous article \cite{MVSHE}, we presented a simple 
statistical model for the description of production 
cross sections of superheavy nuclei in a wide range 
of excitation energies including cold, warm and hot 
fusion. The model assumes that the fusion hindrance, 
observed in cold fusion reactions where only one 
neutron is emitted prior to the formation of evaporation 
residue (ER), can be explained by the competition of 
fusion with fast fission-like process which can 
be identified with quasi-fission. It is not 
obvious what is the role of a traditional saddle configuration, 
used in description of fusion-fission, in quasi-fission.  
Therefore, the scission configuration was chosen as a final state 
in the fission channel. Then the fusion 
probability can be expressed using the level densities 
in compound and scission configurations as 

\begin{equation}
P^{{\rm stat}}_{{\rm fus}} =
\frac{\rho(E^{*}_{{\rm CN}})}{\rho(E^{*}_{{\rm CN}})+\rho(E^{*}_{{\rm sc,eff}})}.
\label{pstat}
\end{equation}

\noindent
The excitation energy in the scission configuration 
is estimated empirically using the systematics of post-scission neutron 
multiplicities. Proportionality of the number of neutrons emitted from the fission fragments 
to the intrinsic excitation energy in the scission configuration is assumed. 
Then the excitation energy in the scission configuration 
can be expressed as 

\begin{equation}
E^{*}_{{\rm sc,eff}} = (\nu_{{\rm n}}^{{\rm s.f.}}(A_{{\rm CN}})
+\Delta\nu_{{\rm n}}(E^{*}_{{\rm CN}}) )E_{\rm n}.
\label{esciss}
\end{equation}

\noindent
The multiplicity of emitted neutrons in the spontaneous fission 
of heavy nuclei $\nu_{{\rm n}}^{{\rm s.f.}}(A_{{\rm CN}})$ 
is approximated by a linear extrapolation 
of the available spontaneous fission neutron multiplicity data
to given $A_{{\rm CN}}$ 

\begin{equation}
\nu_{n}^{{\rm s.f.}}(A_{{\rm CN}}) = 3.316 + 0.0969(A_{{\rm CN}}-250).
\label{mnsf}
\end{equation}

\noindent
An additional increase of the post-scission neutron multiplicity
at a given excitation energy $\Delta\nu_{n}(E^{*}_{{\rm CN}})$ can be expressed
approximately as 

\begin{equation}
\hbox{$\Delta\nu_{{\rm n}}(E^{*}_{{\rm CN}}) = 0.035 E^{*}_{{\rm CN}}$},
\label{mnex}
\end{equation}

\noindent
as follows from the available post-scission neutron multiplicity
data \cite{npost}. A proportionality factor $E_{{\rm n}}$ 
is the amount of intrinsic excitation energy
per emitted neutron. This is a free parameter and it was estimated from 
the systematics of production cross sections of transfermium nuclei 
produced in cold and hot fusion. 

The unhindered fusion cross sections have been calculated 
using one-dimensional WKB approximation with Gaussian barrier 
width distribution \cite{GauBar} implemented into the 
statistical code HIVAP \cite{HIVAP}. Such an approximation proved 
quite successful despite its simplicity. The 
depth of the nuclear potential well is taken as $V_0$=40 MeV, the half-density 
radius as $r_0$=1.11 fm and the diffuseness is set to $d$=0.75 fm. 
The width of barrier distribution ranges from 3 \% for reactions with the doubly magic 
nucleus $^{208}$Pb to 5 \% for reactions with heavy deformed nuclei away from the shell 
closure. 

The survival probabilities were calculated using a conventional statistical calculation. 
The competition of fission vs particle emission was calculated using a modified 
version of HIVAP code \cite{HIVAP} with fission barriers expressed as \cite{CFull} 

\begin{equation}
B_{{\rm f}}(l) = C ( B_{{\rm f}}^{{\rm LD}}(l) + \Delta B_{{\rm f}}^{{\rm Shell}} ).
\label{cfull}
\end{equation}

\noindent
The liquid drop component of the fission barrier  ($B_{{\rm f}}^{{\rm LD}}$)
has been calculated according to the rotating charged liquid drop model
of Cohen-Plasil-Swiatecki
\cite{CO1}. The shell component of the fission 
barrier ($\Delta B_{{\rm f}}^{{\rm Shell}}$) has been approximated by a value 
of the ground state shell correction 
taken from the calculation of Moller \cite{Mo95}. 
Such an approximation for the fission barriers proved 
successful for description of the evaporation residue cross sections 
in the region around neutron shell closure N=126 where the value 
of the parameter $C$ proved to be practically constant for the large 
set of evaporation residues with the values of the ground state shell 
correction ranging from zero up to 8 MeV \cite{CFull}. 

\begin{figure}[t]
\centering
\includegraphics[height=12cm,width=15cm]{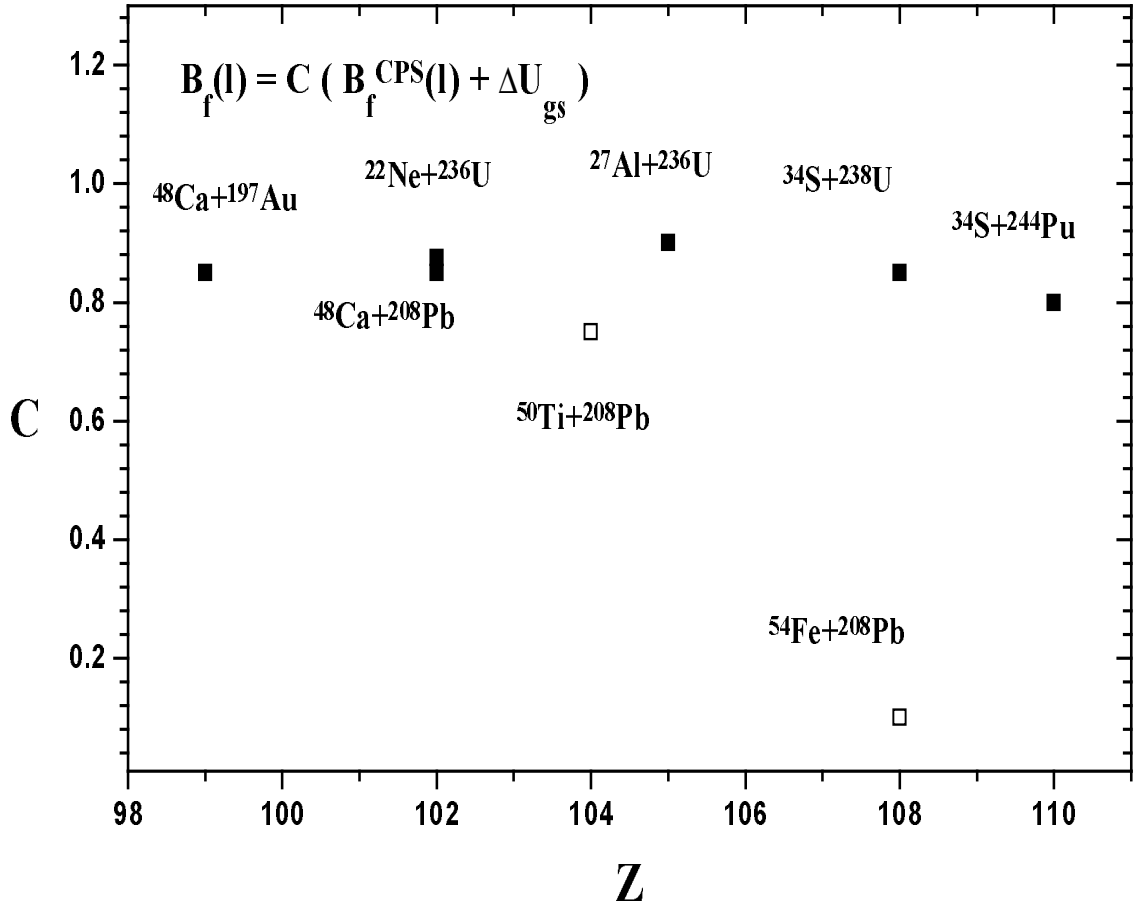}
\caption{ Optimum values of parameter $C$, necessary to reproduce experimental 
cross sections of hot fusion reactions ( solid symbols ), as a function of 
atomic number of residual nuclei. Open symbols - cold fusion reactions. }
\label{figcsyst}
\end{figure}

The shell corrections for transfermium nuclei are expected to be 
within the same range while the liquid drop fission barriers are virtually zero. 
The optimum values of parameter $C$, necessary to reproduce experimental 
cross sections of the evaporation residues with Z$>$100, are given 
in Figure \ref{figcsyst} 
as a function of atomic number. 
One can see that the optimum values of $C$ for the hot fusion reactions remain stable 
within Z = 102 - 110. The value of $C$ = 0.8 - 0.9 is higher when compared 
to N=126 region and this difference could be most probably attributed 
to differences of saddle point configurations in both regions.
 
Unlike for the hot fusion products, the optimum values of the parameter $C$ for the 
cold fusion reactions with $^{208}$Pb target are increasingly falling out of systematics at Z $>$ 104 
what can be attributed to emerging competition of the fusion with quasi-fission. 
Thus, the fusion probabilities for cold fusion were obtained by comparing the measured evaporation residue cross 
sections with those calculated using $C$ from the hot fusion systematics. 
The parametrization $E_{\rm n} = 3.795 + 0.04(A_{{\rm CN}}-260)$ 
for the parameter in the formula \ref{esciss} was obtained and used in further calculation for nuclei 
with Z$>$110. 

The alternative scenario 
of the fusion hindrance originating from tunneling through the 
barrier in the sub-barrier region seems to be contradiction 
with experimental ratios of the cross sections in 1n and 2n 
evaporation channels of reactions with $^{208}$Pb target \cite{Z112Cold} 
which increase from about 0.1 for $^{48}$Ca-beam ( Z=102 ) to 10 for 
$^{58}$Fe-beam ( Z=108 ). Such a situation suggests that even at excitation 
energies corresponding to 1n channel the reaction can not be considered 
of a sub-barrier type. 

Table 1 gives the production cross sections 
of several new superheavy nuclei \cite{Z112Vass,Z114Vass,Z114GNS,Z116GNS}, 
reported since our initial article \cite{MVSHE} was published, compared to the maximum 
evaporation residue cross sections in xn channels estimated in \cite{MVSHE}. 
No angular momentum dependence for description of the scission point was assumed and 
the cross sections were evaluated in the maxima 
of the excitation functions obtained from statistical 
calculations with no fusion hindrance assumed. 
One can see that the calculation predicted production cross 
sections rather well for the reaction $^{48}$Ca+$^{238}$U. For heavier systems 
the estimated cross sections are lower by up to one order of magnitude since calculation 
exhibits systematic shift in the dominating xn evaporation residue channel 
toward higher number of emitted neutrons. 
The discrepancy observed can be attributed to the simplification 
used in the initial calculation where the unhindered maximum production 
cross sections obtained using HIVAP code were multiplied by the 
fusion probabilities with no angular momentum dependence assumed. 
Recently, the calculation was corrected \cite{MVPrgRep} by implementing 
an angular momentum into the description of the compound nucleus and scission 
configuration and by introducing the fusion probability calculation 
for each partial wave into HIVAP code. The moment of inertia of symmetric touching 
rigid spheres was used for the scission configuration. 
An improved version of HIVAP code uses the fusion probability for each partial wave 
as a multiplication factor to the unhindered fusion cross section. 
This allows to obtain more realistic shapes of excitation functions for 
evaporation residue channels. 

In Table 2 are again given production cross sections of the recently synthesized superheavy nuclei, compared to the 
results of improved calculation \cite{MVPrgRep}. The production cross sections track quite well 
with the reported ones. New calculation reproduces reasonably well not only the 
absolute values but also the positions of the maxima and thus promises possibility for further estimates. 
Concerning the recently reported \cite{Z118BGS} ( and more recently corrected \cite{Z118Ret} ) 
experimental results from the reaction $^{86}$Kr+$^{208}$Pb, the calculation ( as published in \cite{MVPrgRep} ) 
lead to estimated production cross section for 1n channel of approximately 10$^{-4}$ pb. 
Such a value was in contradiction with initial experimental cross section value 2.2 pb \cite{Z118BGS} 
but it is consistent with the corrected experimental results. As stated above, 
the parametrization of $E_{{\rm n}}$ used was obtained using data from cold fusion only 
and thus the estimated cross section for cold fusion is practically just an extrapolation 
of cross section trend. In any case, significant success of the extrapolation when 
used for hot fusion reactions virtually justifies its validity also for cold fusion 
reactions. Therefore, the cold fusion reactions do not seem to offer much promise 
for further progress in the synthesis of superheavy nuclei.  

In Table 3 are given predictions for several 
reactions which may lead to the synthesis of even 
heavier nuclei. An improved calculation \cite{MVPrgRep} was used 
in this case. Only reactions of stable beams with stable 
or long-lived targets have been taken into account. 
The reactions $^{48}$Ca+$^{249}$Cf and $^{58}$Fe+$^{238}$U give   
promise for the synthesis of the isotope $^{292,293}$118 on the cross section 
level of 0.1-0.2  pb which seems to be an experimental limit 
for the foreseeable future. Compared to the system $^{58}$Fe+$^{238}$U, 
the choice of heavier projectile $^{64}$Ni or target $^{244}$Pu leads 
to the drop of cross section by one and half orders of magnitude. 
It is necessary to note that the quality of the estimate directly 
depends on the prediction of the masses and ground state shell 
corrections \cite{Mo95} used in the calculation. The results 
given in Table 2 suggest that the masses and ground state shell corrections 
used are quite realistic. Nevertheless, any discrepancies in further 
extrapolation will affect the cross section estimates significantly.  

\section*{Capture cross sections}

\begin{figure}[t]
\centering
\includegraphics[width=15cm]{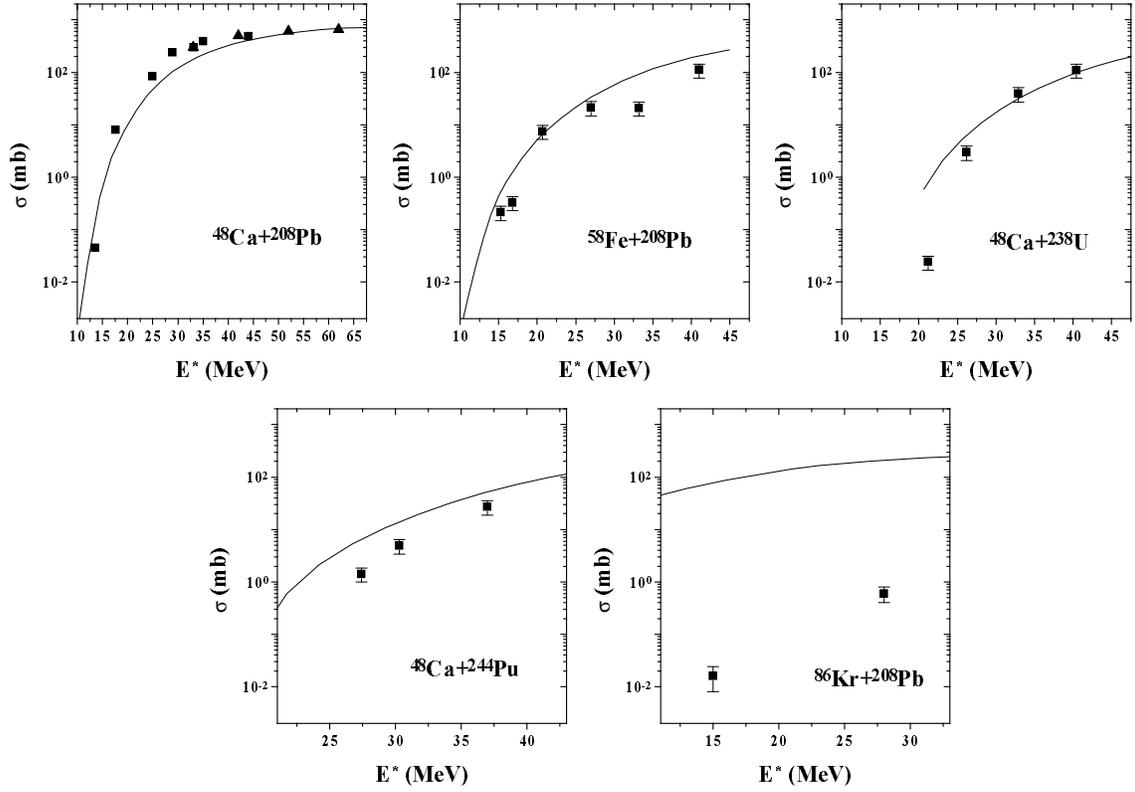}
\caption{ Experimental capture cross sections \cite{Itkis} ( symbols ) 
and the fusion cross sections calculated using WKB approximation with 
Gaussian barrier distribution \cite{GauBar} ( lines ). Data from five 
different reactions are presented. Width of barrier distribution is assumed 
3\% for $^{208}$Pb target and 5\% for $^{238}$U and $^{244}$Pu targets. }
\label{figcpt1}
\end{figure}

\begin{figure}[t]
\centering
\includegraphics[width=14cm]{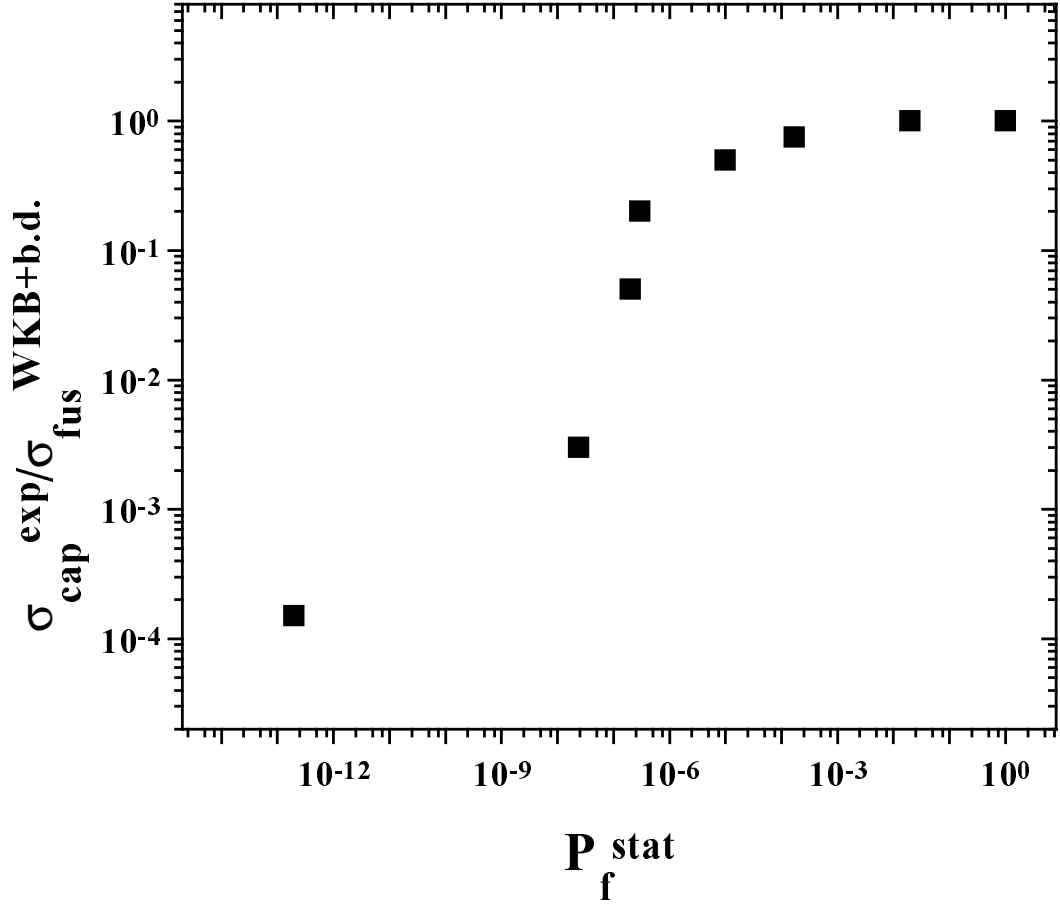}
\caption{ Ratio of the experimental capture cross sections \cite{Itkis} 
to the fusion cross sections calculated using WKB approximation with 
Gaussian barrier distribution \cite{GauBar} ( symbols ) plotted as 
a function of the calculated fusion probability \cite{MVSHE,MVPrgRep}. 
The data points from Fig. \ref{figcpt1} are used. }
\label{figcpt2}
\end{figure}

\begin{figure}[t]
\centering
\includegraphics[width=14cm]{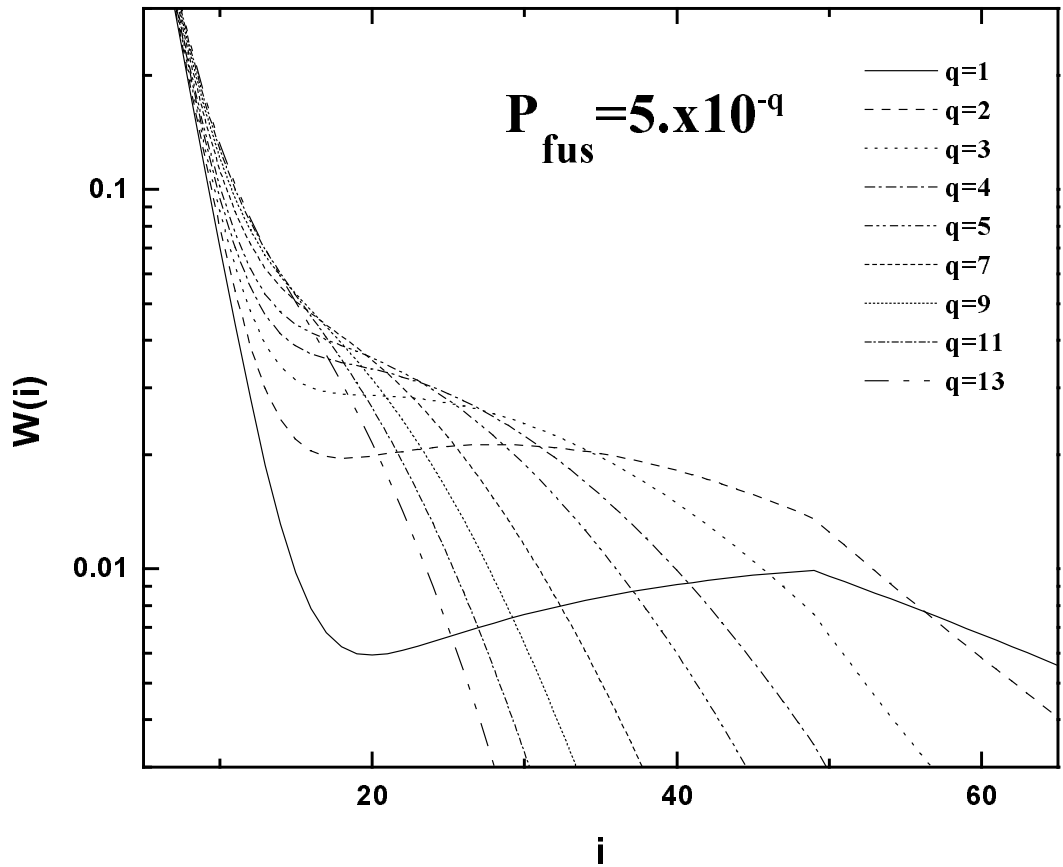}
\caption{ "Toy model" description of the fusion-fission vs quasi-fission 
competition during the dynamical evolution of the system. The exit channel 
probability density W(i) is plotted as a function of step number i for 
several values of the resulting fusion probability $P_{{\rm fus}}$. 
For details see text.}
\label{figcpt5}
\end{figure}

In order to understand the competition of the fusion and quasi-fission it is 
of great interest to investigate also the measured cross sections 
of the fusion-fission and quasi-fission. Such an analysis was performed 
on the data on measured capture cross sections \cite{Itkis} ( defined in the experiment as the cross 
section of the fission events with the total kinetic energy and fragment 
masses outside of the quasi-elastic/deep-inelastic region of the 
TKE vs mass matrix ). The experimental setup 
used was optimized to detect fusion-fission for each specific reactions. 
A comparison of the experimentally determined capture cross sections 
\cite{Itkis} to the calculated unhindered fusion cross sections \cite{GauBar} 
and fusion probabilities \cite{MVSHE,MVPrgRep} is presented 
in Figures \ref{figcpt1},\ref{figcpt2}. 
As one can see from the Fig. \ref{figcpt1} the calculated fusion cross 
sections track very well with the measured capture cross sections for 
the reactions $^{48}$Ca+$^{208}$Pb and $^{58}$Ca+$^{208}$Pb. For the 
heavier systems the measured capture cross sections become smaller than the 
calculated fusion cross sections. Such an effect appears to increase 
with decrease of the excitation energy of the compound nucleus. One can assume 
that such a discrepancy can be related to decrease of the fusion 
probability for the heavier systems. Such an assumption is examined 
in the Fig. \ref{figcpt2} where the ratio of the measured capture 
cross section to the calculated unhindered fusion cross sections is represented 
as a function of the fusion probability calculated as in \cite{MVPrgRep}. One can 
observe a surprising abrupt disappearance of the measured capture 
cross section at fusion probabilities below 10$^{-6}$. 
Such an abrupt disappearance of the measured capture cross section when compared to the calculated 
fusion cross section seems to be rather global and it may indicate a dramatic change of the properties of reaction 
products due to different dynamical evolution. A possibility to explain the trend 
qualitatively is presented in Figure \ref{figcpt5} in the framework of "toy model" mimicking 
a competition of multi-step dynamical evolution toward fusion with 
a possibility of irreversible exit into the quasi-fission channel at each step. 
Fusion probability is treated as a product of N elementary sub-probabilities P(i) corresponding 
to elementary steps of the evolution toward fusion. 
The probability for the first step P(1) is assumed one, later the probability 
decreases linearly until it reaches minimum halfway toward fusion, 
then the probability starts to increase linearly and the probability of the 
last step is again assumed one. At each step, the quantity 1-P(i) can be 
considered the probability of exit into quasi-fission channel. 
The resulting exit channel probability density of fusion--quasi-fission 
competition with 100 steps is superimposed onto the exit channel probability 
density of another process with the exit channel probability density quickly 
exponentially decreasing with step number. The latter process is considered 
10 times more frequent. Such a procedure can simulate an interplay with 
the quasi-elastic/deep-inelastic reactions occurring at the partial waves 
close to the grazing angular momentum ( and thus with higher cross section ). 
As one can see with decreasing fusion probability the exit channel probability 
densities of two processes increasingly overlap and at some point can not 
be decomposed anymore. This can be a qualitative explanation for the situation 
in Fig. \ref{figcpt2} where the measured capture cross-section initially tracks with the 
calculated fusion cross section but at some point this correspondence 
disrupts abruptly. In the realistic process leading to either fusion or quasi-fission, 
the concentration of the probability density at the early stage of dynamical evolution may lead to kinematic properties 
of the fission fragments very different from the fusion-fission. Such a fragments can become  
undetectable using a given experimental setup optimized for detection of the fusion-fission products.  
In any case the disappearance of the measured capture cross sections in a given case 
can be understood as a signature of the interconnection of the fusion and the quasi-fission processes 
within the concept of their competition during the multi-step dynamical evolution 
of the system.  

\section*{Symmetric systems}

Of great interest for the future prospects of synthesis of superheavy nuclei is 
the understanding of reaction dynamics in the case where both projectile and target 
are of comparable size. In order to investigate a possible fusion hindrance originating from 
increasing symmetry of the projectile-target system we compared the calculated evaporation residue 
cross section in the four reactions leading to compound nucleus $^{246}$Fm 
to the experimental cross sections from the work of Gaeggeler et al. \cite{Gaegg}. 
The calculation used was identical to \cite{MVPrgRep}. The result is presented in the Table 4. 
When looking at the results and taking into account the systematics in Fig. \ref{figcsyst} 
where fusion hindrance occurs for cold fusion of compound nuclei with Z$>$104, 
one can assume that there appear to exist additional 
fusion hindrances which emerge with increasing symmetry of the reaction.

\begin{figure}[t]
\centering
\includegraphics[width=14cm]{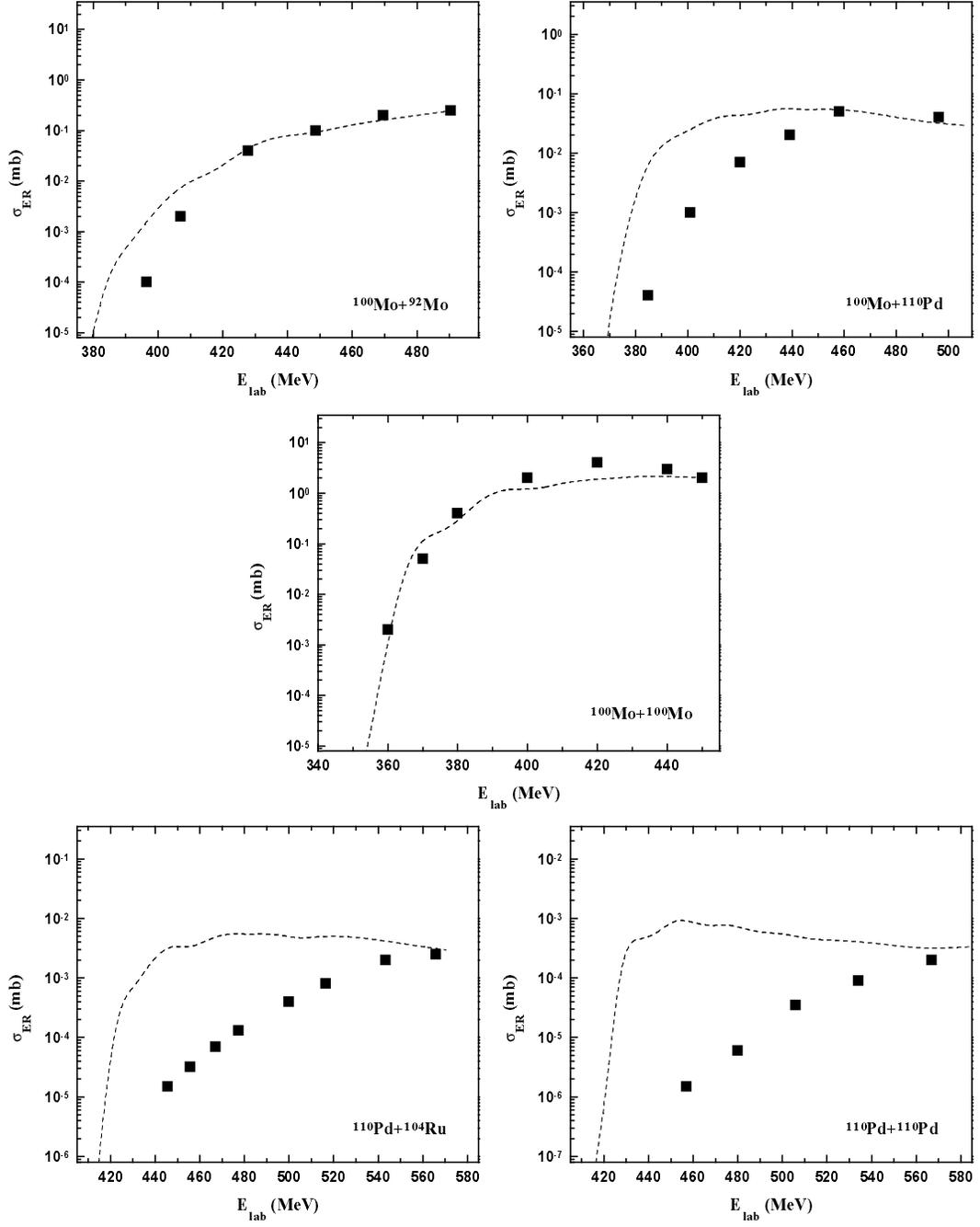}
\caption{ Comparison of the measured \cite{Morawek,Quint} 
and calculated evaporation residue cross sections 
for several symmetric systems. }
\label{ercssym}
\end{figure}

In order to understand a possible nature of such hindrances
we carried out an analysis of the data in the Pb-U region \cite{Morawek,Quint}. 
Using the fusion model with WKB and Gaussian
barrier distribution \cite{GauBar} and fission channel parameters from the systematics
for given region \cite{CFull} ( $C$ $\approx$ 0.65 ) we observe an interesting behavior ( see Fig. \ref{ercssym} ).
For the reaction $^{100}$Mo+$^{100}$Mo the evaporation residue cross section is described well. 
In the transition to $^{110}$Pd+$^{110}$Pd there is an increasing hindrance at low excitation energies.
The hindrance factor seems to increase with decreasing excitation energy.
To some surprise, the same effect can be seen also in the transition from $^{100}$Mo+$^{100}$Mo
to $^{100}$Mo+$^{92}$Mo ( lighter system but with higher fissility ). 
Also of interest is the fact that experimental cross section data for Pd+Ru and
Pd+Pd systems are only in the region above calculated fusion barrier where calculated fusion cross remain
stable but disappear in the sub-barrier region where calculated cross sections start to drop quickly.

\begin{figure}[t]
\centering
\includegraphics[width=14cm]{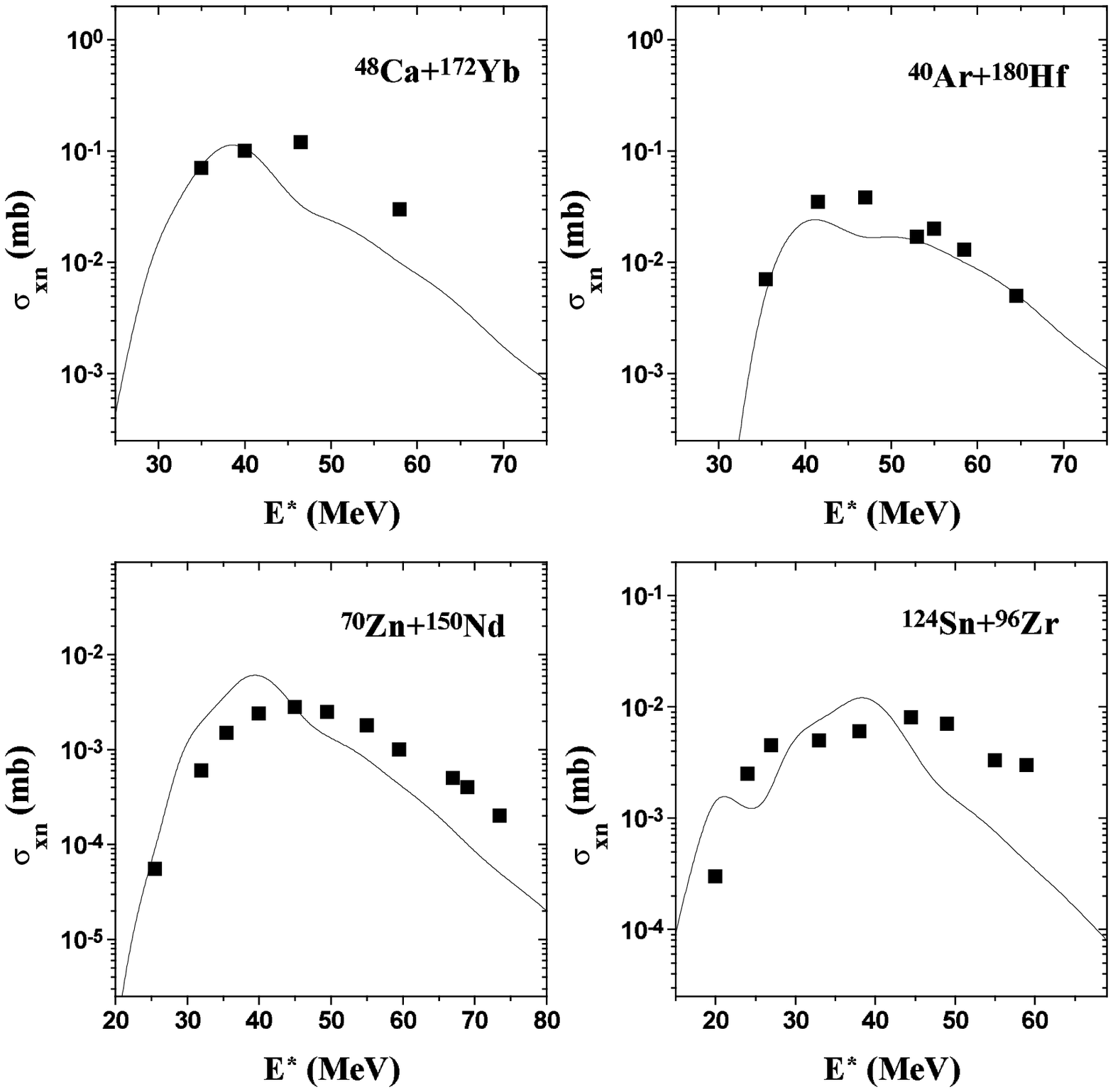}
\caption{ Comparison of the measured \cite{Vermeulen,Sahm1,Peter} 
and calculated xn evaporation residue cross sections 
for several systems leading to the compound nucleus $^{220}$Th. }
\label{xncssym}
\end{figure}

In order to test the fusion cross section model, the measured 
and calculated xn evaporation cross section for four systems leading to compound nucleus $^{220}$Th 
( $^{40}$Ar+$^{180}$Hf \cite{Vermeulen}, $^{124}$Sn+$^{96}$Zr \cite{Sahm1}, $^{48}$Ca+$^{172}$Yt and $^{70}$Zn+$^{150}$Gd \cite{Peter} )
are given in the Fig. \ref{xncssym}. 
The calculations have been performed using HIVAP code. The barrier distribution widths \cite{GauBar} 
used comply to the usual prescription ( 5 \% for Ar+Hf, Ca+Yt and Zn+Gd
since ( heavy ) target is deformed and 4 \% for Sn+Zr since
( heavier ) projectile is close to spherical ). The shapes
of xn excitation functions are reproduced reasonably well, especially  
the ascending/barrier part and the maximum of xn excitation functions are
reproduced acceptably. The fission barrier scaling parameter $C$ \cite{CFull} was equal for Ar+Hf 
and Ca+Yt ( $C$=0.67 ) and Zn+Gd and Sn+Zr ( $C$=0.61 ). 
The discrepancy in $C$ is not fully compliant with the
concept of compound nucleus, since it should be the same 
in all cases. Most probably it is caused 
by the irregularities in alpha-emission where especially in the symmetric systems 
the memory of the entrance channel ( e.g. deformation ) may lead to enhanced 
emission of the alpha-particles and thus to reduction of xn cross sections. 
Experimental alpha-particle emission spectra \cite{AlVaz} suggest alpha emission barriers of about
90 \% of the alpha-particle fusion barrier what is also used in calculations 
but such a prescription is rather simplistic and may not account 
for dynamical effects in symmetric reactions. Apart from entrance channel memory, 
an admixture from incomplete fusion channels
with emission of alpha-particle is also possible. More detailed data
will be necessary for complete understanding of the phenomena. 
In any case, the description of the fusion barrier by the approximation 
employed can be considered adequate.

\begin{figure}[t]
\centering
\includegraphics[width=14cm]{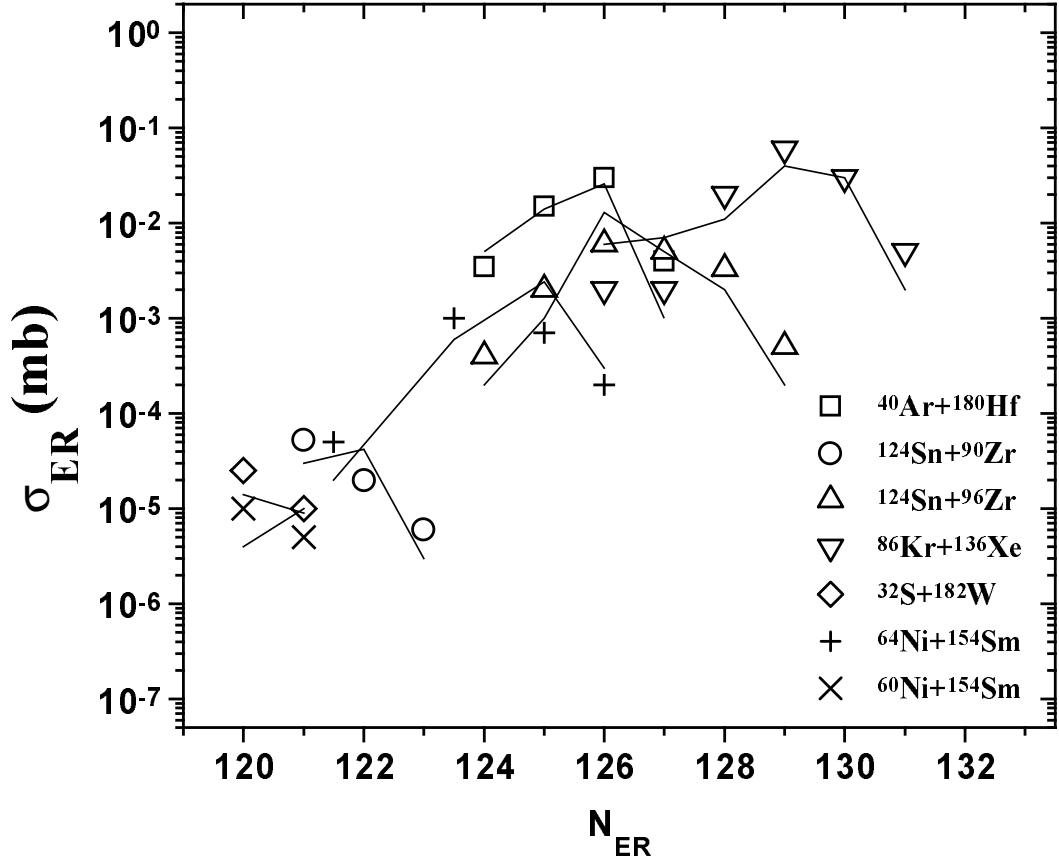}
\caption{ Comparison of the measured 
\cite{Vermeulen,Sahm1,Sahm2,Mitsuoka1,Mitsuoka2,Sagaidak}
and calculated maximum xn evaporation residue cross sections 
for several systems leading to various Th compound nuclei. }
\label{thxncs}
\end{figure}

Further comparisons of the calculated and experimental evaporation residue 
cross sections for reactions leading to various Th compound nuclei 
are given in Fig. \ref{thxncs}. The maximum cross sections for various xn evaporation 
channels are considered. For the compound system $^{214}$Th where one can see a strong hindrance 
for reaction $^{110}$Pd+$^{104}$Ru ( see Fig. \ref{ercssym} ) the same can not be concluded 
for the reaction $^{124}$Sn+$^{90}$Zr \cite{Sahm2}. 
Also, for Th compound nuclei ranging from $^{214}$Th to $^{222}$Th 
no hindrance can be observed for reactions including $^{32}$S+$^{182}$W \cite{Mitsuoka1}, 
$^{60}$Ni+$^{154}$Sm \cite{Mitsuoka1}, $^{64}$Ni+$^{154}$Sm \cite{Mitsuoka2} and $^{86}$Kr+$^{136}$Xe \cite{Sagaidak}.
For $^{86}$Kr+$^{136}$Xe data the xn cross sections are practically constant from 1n to 6n channel
what is in conflict with extra-push theory \cite{XPush}. As in the previous 
case the fission barrier scaling parameter $C$ varied from 0.6 to 0.67 and emission barriers 
were 10 \% lower than fusion barriers for a given light charged particle. 
The widths of the fusion barrier distribution were consistent to above prescription. 
As one can see from Figs. \ref{xncssym} and \ref{thxncs}, statistical model calculation 
with fission barriers compliant to the formula \ref{cfull} ( giving equally good description 
for nuclei with and without strong g.s. shell corrections \cite{CFull} ) and with fusion cross section 
calculated using one-dimensional WKB approximation with fusion barrier distribution 
provides very consistent description of the evaporation residue cross sections virtually without 
using free parameters. No fusion hindrance can be observed for a wide range of 
compound nuclei. Thus one can conclude that the fusion hindrance in Th-region 
takes place only for the reactions leading to highly fissile compound nuclei with 
projectile-target asymmetry in narrow region close to zero.

When looking for an explanation of the above behavior one can turn attention to the 
properties of the fission fragments in the given region. 
Recent studies of low energy fission in Ac-U region \cite{FissKHS,FissTx}  show that there 
is a systematic transition from asymmetric to symmetric fission around the mass 222-226. The Th compound 
nuclei studied above all fall into the region with symmetric fission mode. Thus, one can assume that
an additional fission hindrance appears when the asymmetry of fusion
channel is close to the asymmetry of fission channel. There, one can assume that
immediate fission is highly favored dynamically over the long evolution toward
fusion. For the heavier nuclei with masses above 226 the dominant fission mode
at low excitation energies is the mode where one fragment  ( heavier one for
lighter nuclei and lighter one for very heavy nuclei ) is of the mass approximately 132 and the mass of
the other fragment increases linearly with the mass of fissioning system \cite{Itkis}. 
The reaction $^{136}$Xe+$^{110}$Pd studied by Gaeggeler et al. \cite{Gaegg} is virtually 
an inverse fission and a dynamical fusion hindrance can be understood there. 
The reactions $^{76}$Ge+$^{170}$Kr, $^{86}$Kr+$^{160}$Gd are far away from the main fission mode 
but still match the super-asymmetric mode ( with the maximum yiels of light fragment positioned 
around $^{82}$Ge ) which is usually necessary to reproduce the 
experimental mass distributions \cite{Rubch}. Thus, the knowledge of fission modes in
transfermium region seems to be an essential information for the study of fusion probability. 
An interesting test for such an assumption would be a reaction $^{132}$Sn+$^{96}$Zr leading to compound nucleus
$^{228}$Th which fissions asymmetrically and thus it would be an inverse fission
again and hindrance factors should appear. The non-hindered cross sections
can be expected in mb region so already a relatively moderate beam of $^{132}$Sn 
may be sufficient to show discrepancy. The use of radioactive beam is essential in this case
since no symmetric combination of the stable beam and target appears to
reach Th isotopes beyond 222. 

\section*{Summary and conclusions}

In summary, the possibilities for synthesis of new 
superheavy elements using stable or long-lived projectiles 
and targets seem to be rather restricted. The parametrization 
of model parameters able to reproduce existing experimental 
results predicts a possibility to synthesize isotopes of element 118 
in hot fusion reactions at the cross section level 0.1 pb. 
Concerning the nature of the process, the analysis of the 
measured cross sections suggests that the competition of 
fusion and quasi-fission is a multi-step dynamical process and that 
the low fusion probability is consistent with the fast re-separation 
of the reacting system even at low partial waves. 
For symmetric systems where the asymmetry of the projectile-target combination 
approaches the asymmetry of the fission channel an additional 
fusion hindrance caused by dynamical dominance of immediate  
re-separation into the fission channel over long evolution 
toward complete fusion seems to take place. Such a dynamical hindrance can 
strongly reduce the possible pathways toward superheavy elements. 
A knowledge on fission fragment asymmetry seems to be essential 
for further studies of synthesis of superheavy elements. 

The author would like to thank S. Hofmann, A.V. Yeremin, J. Peter, R. Smolanczuk, 
G. Chubarian, E.M. Kozulin, W. Loveland and G.A. Souliotis for fruitful discussions. 
Furthermore, the author would like to thank Yu.Ts. Oganessian and M.G. Itkis for their 
interest to this work. This work was supported through grant VEGA-2/1132/21.

\newpage

\begin{center}
\begin{table}[tbph]
\caption{ Comparison of several recently reported production cross 
sections of elements with Z$>$110 \cite{Z112Vass,Z114Vass,Z114GNS,Z116GNS} to the predictions published in \cite{MVSHE}.}

\vskip 0.4cm
{\centering \begin{tabular}{ccccc}
\hline
Reaction &\multicolumn{2}{c}{Experiment}&\multicolumn{2}{c}{Calculation}  \\ 
         & ER & $\sigma_{ER}$ & ER & $\sigma_{ER}$ \\ \hline
$^{48}$Ca+$^{238}$U & $^{283}$112 & 5 pb & $^{283}$112 &  1.5 pb  \\
$^{48}$Ca+$^{242}$Pu & $^{287}$114 & 2.5 pb & $^{286}$114 &  0.25 pb  \\
$^{48}$Ca+$^{244}$Pu & $^{288}$114 & 0.7 pb & $^{287}$114 & 0.1 pb  \\
$^{48}$Ca+$^{248}$Cm & $^{292}$116 & 0.3 pb & $^{291}$116 & 0.01 pb  \\ \hline
\end{tabular} \par}
\end{table}

\end{center}


\begin{center}
\begin{table}[tbph]
\caption{ Comparison of recently reported production cross 
sections of elements with Z$>$110 \cite{Z112Vass,Z114Vass,Z114GNS,Z116GNS} 
to the results of the improved calculations \cite{MVPrgRep}. }

\vskip 0.4cm
{\centering 
\begin{tabular}{ccccc}
\hline
Reaction & E$_{lab} $ & ER & \multicolumn{2}{c}{$\sigma_{ER}$ [pb]} \\
 & [MeV]  &  & Exp. & Calc. \\ \hline
$^{48}$Ca+$^{238}$U & 231 & $^{283}$112 & 5 & 4  \\
$^{48}$Ca+$^{238}$U & 238 & $^{282}$112 & $\le$ 7 & 8  \\
$^{48}$Ca+$^{242}$Pu & 235 & $^{287}$114 & 2.5 & 1.5  \\
$^{48}$Ca+$^{244}$Pu & 236 & $^{288}$114 & 0.7 & 2.0  \\
$^{48}$Ca+$^{248}$Cm & 240 & $^{292}$116 & 0.3 & 0.1  \\ \hline
\end{tabular}
\par}
\end{table}

\end{center}


\begin{center}
\begin{table}[tbph]
\caption{ Predictions of production cross 
sections of elements with Z$>$116 calculated using 
improved calculation \cite{MVPrgRep}. Reactions of stable beams with stable 
or long-lived targets have been taken into account. }

\vskip 0.4cm
{\centering 
\begin{tabular}{cccc}
\hline
Reaction & E$^{*}$, MeV  & ER & $\sigma_{ER}$(calc) \\ \hline
$^{48}$Ca+$^{249}$Cf & 47 & $^{293}$118 & 0.1 pb \\
$^{48}$Ca+$^{249}$Cf & 52 & $^{292}$118 & 0.25 pb \\
$^{48}$Ca+$^{252}$Cf & 46 & $^{296}$118 & 0.02 pb  \\
$^{48}$Ca+$^{252}$Cf & 53 & $^{295}$118 & 0.03 pb  \\
$^{58}$Fe+$^{238}$U & 48 & $^{292}$118 & 0.2 pb  \\
$^{58}$Fe+$^{244}$Pu & 56 & $^{297}$120 & 0.007 pb  \\
$^{64}$Ni+$^{238}$U & 56 & $^{297}$120 & 0.007 pb  \\ \hline
\end{tabular}
\par}
\end{table}

\end{center}

\newpage

\begin{center}
\begin{table}[tbph]
\caption{ Comparison of the calculated evaporation residue 
cross section in the four reactions leading to compound nucleus $^{246}$Fm 
to the experimental cross sections from the work of Gaeggeler et al. \cite{Gaegg}. }

\vskip 0.4cm
{\centering 
\begin{tabular}{ccc}
\hline
Reaction   &     $\sigma_{2n}$(exp)   &      $\sigma_{2n}$(calc) \\ \hline
$^{40}$Ar+$^{206}$Pb     &   3 nb               &     1 nb \\ 
$^{76}$Ge+$^{170}$Er     &   1 nb                &    19 nb \\ 
$^{86}$Kr+$^{160}$Gd     &   $<$ 0.3 nb             &   26 nb \\ 
$^{136}$Xe+$^{110}$Pd    &   $<$ 0.2 nb              &  40 nb \\ \hline
\end{tabular}
\par}
\end{table}

\end{center}

\newpage

\end{document}